\documentclass[preprint,prc,showpacs,preprintnumbers,amsmath,amssymb]{revtex4}

\usepackage{graphicx}
\usepackage{dcolumn}
\usepackage{bm}
\usepackage{multirow}
\usepackage{color}

\newcommand{\br}{\mbox{\boldmath$r$}}

\begin{document}

\title{Magnetic moments of $^{33}$Mg in time-odd relativistic mean field approach}

 \author{Jian~Li~}
 \affiliation{School of Phyics, State Key Laboratory of Nuclear Physics
   and Technology, Peking University, Beijing 100871, China}
 \author{Ying~Zhang~}
 \affiliation{School of Phyics, State Key Laboratory of Nuclear Physics
   and Technology, Peking University, Beijing 100871, China}
 \author{J. M. Yao}%
 \affiliation{School of Phyics, State Key Laboratory of Nuclear Physics
   and Technology, Peking University, Beijing 100871, China}
 \author{J. Meng}
 \email{mengj@pku.edu.cn}%
  \affiliation{School of Phyics, State Key Laboratory of Nuclear Physics
   and Technology, Peking University, Beijing 100871, China}
 \affiliation{School of Physics and Nuclear Energy Engineering, Beihang University, Beijing 100191, China}

\begin{abstract}
The configuration-fixed deformation constrained relativistic mean
field approach with time-odd component has been applied to
investigate the ground-state properties of $^{33}$Mg with effective
interaction PK1. The ground state of $^{33}$Mg has been found to be
prolate deformed, $\beta_2=0.23$, with the odd neutron in $1/2[330]$
orbital and the energy $-251.85$ MeV which is close to the data
$-252.06$ MeV. The magnetic moment $-\,0.9134\,\mu_\mathrm{N}$ is
obtained with the effective electromagnetic current which well
reproduces the data $-\,0.7456\,\mu_\mathrm{N}$ self-consistently
without introducing any parameter. The energy splittings of time
reversal conjugate states, the neutron current, the energy
contribution from the nuclear magnetic potential, and the effect of
core polarization are discussed in detail.
\end{abstract}

 \maketitle
\section{Introduction}

Nuclear magnetic moment, as one of the most important physical
observables, provides key information to understand nuclear
structure, and has attracted the attention of nuclear physicists
since the early days~\cite{Blin-Stoyle1956,Arima1984,Castel1990}.
Apart from the magnetic moments of stable nuclei~\cite{Stone2005},
it is now even possible to measure the nuclear magnetic moments of
many short-lived nuclei far from the stability line with high
precision~\cite{Neyens2003} with the development of the radioactive
ion beam (RIB) technique.

Recently the nuclear magnetic moments of $^{33}$Mg has become a hot
topic due to the following reasons: 1) it is a neutron-rich nucleus
close to the so-called ``island of inversion"~\cite{Warburton1990};
2) different spins and configurations for the ground state of
$^{33}$Mg are assigned in a series of
experiments~\cite{Nummela2001,Pritychenko2002,Elekes2006,Tripathi2008}.
In order to remove the confusion, the spin and magnetic moment for
the ground state in $^{33}$Mg have been directly measured in
Ref.~\cite{Yordanov2007} with $I=3/2$ and
$\mu=-0.7456(5)\mu_\mathrm{N}$, which becomes a test for various
theoretical approaches. In shell-model, the magnetic moment of the
ground-state in $^{33}$Mg, can be reproduced only in the model space
with $2p-2h$ configuration~\cite{Yordanov2007}. With the assignment
of configuration in Ref.~\cite{Tripathi2008}, the simple Additivity
Rules~\cite{Lawson1980} can only account for half of the
experimental magnetic moment.

For the last two decades, the relativistic mean field (RMF)
theory~\cite{SEROT1986} has achieved great success in describing
many phenomena in the nuclei near and far from the line of
$\beta$-stability~\cite{Ring1996,Vretenar2005,Meng2006}. In the
widely used version of RMF models, only the time-even component of
vector meson fields are taken into account. In odd-A or odd-odd
nuclei, however, the baryon current due to the unpaired valence
nucleons will lead to the time-odd components of vector fields,
i.e., the time-odd fields. The time-odd fields together with the
corresponding core polarization will modify the nuclear current,
single-particle Dirac spinor, and magnetic moments, etc. The
importance of the time-odd fields has been demonstrated in the
successful descriptions of nuclear magnetic
moments~\cite{Hofmann1988,Furnstahl1989,Yao2006}, moment of inertia
for identical superdeformed bands~\cite{Konig1993}, and $M 1$
transition rates in magnetic rotation~\cite{Madokoro2000,Peng2008a},
etc.

In this paper, the ground-state properties of $^{33}$Mg including
the binding energy and magnetic moment will be investigated in
axially deformed RMF approach with time-odd components of vector
meson field~\cite{Li2009}. In order to examine the magnetic moments
for different valence nucleon configurations, both adiabatic and
configuration-fixed deformation constrained
calculation~\cite{Meng2006a} will be performed.


\section{Theoretical Framework}
The basic ansatz of RMF theory is a lagrangian
density~\cite{SEROT1986,Reinhard1989,Ring1996,Meng2006} where
nucleons are described as Dirac particles which interact via the
exchange of various mesons and photon. The lagrangian density is
written in the form
\begin{eqnarray}
   \label{lagrangian}
   {\cal L}
     &=&\overline{\psi}[i{\gamma^\mu}{\partial_\mu}-M
         -{g_\sigma}\sigma- g_\omega\gamma^\mu\omega_\mu
         - g_\rho \gamma^\mu {\vec{\tau}}\cdot{\vec{\rho}}_\mu\nonumber\\
     & & - e\gamma^\mu\frac {1-\tau_3}{2} A_\mu ]\psi+\frac{1}{2}\partial^\mu\sigma\partial_\mu\sigma
       -\frac{1}{2}m_\sigma^2\sigma^2- \frac{1}{3}g_2\sigma^3\nonumber\\
     &&-\frac{1}{4}g_3\sigma^4-\frac{1}{4}\Omega^{\mu\nu}\Omega_{\mu\nu}+\frac{1}{2}m_\omega^2\omega^\mu\omega_\mu
       +\frac{1}{4}c_3\left(\omega^\mu\omega_\mu\right)^2\nonumber\\
     &&-\frac{1}{4}\vec{R}^{\mu\nu}\cdot\vec{R}_{\mu\nu}
       +\frac{1}{2}m_\rho^2\vec{\rho}^\mu\cdot\vec{\rho}_\mu-\frac{1}{4}F^{\mu\nu}F_{\mu\nu}.
\end{eqnarray}
The meson fields contain the isoscalar $\sigma$ meson,
isoscalar-vector $\omega$ meson and the isovector-vector $\rho$
meson. $M$ and $m_i(g_i)~(i = \sigma,\omega,\rho)$ are the masses
(coupling constants) of the nucleon and the mesons respectively and
\begin{subequations}
\begin{eqnarray}
 \Omega^{\mu\nu} &=& \partial^\mu\omega^\nu-\partial^\nu\omega^\mu \\
 \vec{R}^{\mu\nu} &=& \partial^\mu\vec{\rho}^\nu-\partial^\nu\vec{\rho}^\mu\\
 F^{\mu\nu}&=& \partial^\mu A^\nu-\partial^\nu A^\mu
\end{eqnarray}
\end{subequations} are the field tensors of the vector mesons and the
electromagnetic field. In this paper, we adopt the arrows to
indicate vectors in isospin space and bold type for the space
vectors.

For even-even nuclei with time reversal symmetry, the time-odd
components of vector mesons and photon fields do not contribute to
the energy functional. In odd-A nuclei, the odd nucleon breaks the
time-reversal invariance, and time-odd fields give rise to the
nuclear magnetic potential, i.e., the time-odd component of the
vector potential. Then the equation of motion for nucleon can be
obtained as:
  \begin{equation}
   \label{Dirac}
    \{ \bm{\alpha}\cdot[-i\mathbf{\nabla}-\mbox{\boldmath$V$}(\br)]+V_0(\br)
     +  \beta [ M+S({\br}) ] \} \psi_{i}
     = \varepsilon_i\psi_i,
  \end{equation}
with the scalar potential $S(\br)=g_\sigma\sigma(\br)$,
the time-like component of vector potential $V_0(\br) = g_{\omega}\omega_0 (\br) + g_{\rho}\tau_{3} \rho_0 (\br)
+ e{1-\tau_{3} \over 2} A_0({\br})$, and the
time-odd component of vector potential $\mbox{\boldmath$V$}(\br) = g_{\omega}\bm{\omega}
({\br})$,
where $\bm{\rho}(\br)$ and $\mbox{\boldmath$A$}(\br)$ are neglected since they turn out to be small
compared with $\bm{\omega}(\br)$ field in light nuclei~\cite{Hofmann1988}.

The Klein-Gordon equations for $\sigma$, time-like components of vector
mesons fields \emph{$\omega_0$, $\rho_0$} and electromagnetic fields \emph{$A_0$} are the
same as in the Ref~\cite{Meng2006}.
The time-odd component of $\omega$ meson is determined by
\begin{equation}
   \label{KG}
    \{ -\Delta + m_{\omega}^{2} \} \mbox{\boldmath$\omega$}
    =g_\omega \mbox{\boldmath{$j$}}_\mathrm{B} -
    c_3\omega^\nu\omega_\nu\mbox{\boldmath$\omega$},
\end{equation}
with the baryon current
$\mbox{\boldmath{$j$}}_\mathrm{B}=\sum\limits_{i}
n_i\bar\psi_{i}\mbox{\boldmath$\gamma$}\psi_{i}$. The summation is
confined to the particle states with positive energies in the no-sea
approximation. As the pair correlation is neglected here, the
occupation numbers $n_i$ take the value one (zero) for the states
below (above) the Fermi surface. Restricting to an axially symmetric
representation here, only azimuthal baryon currents
$j_\mathrm{B}^\mathrm{\varphi}(z,r_\bot )$ on circular lines around
the symmetry axis are non-zero~\cite{Hofmann1988}, i.e., only the
azimuthal component of the $\omega$ vector field exists.

The total energy of the system, including the time-odd fields, is,
\begin{equation}
     E =\ E_{\mathrm{part}}+E_\mathrm{\sigma} + E_\mathrm{\omega} + E_\mathrm{\rho} +
        E_{\mathrm{c}} + E_{\mathrm{c.m.}}
\end{equation}
with
\begin{eqnarray}\label{tomega}
    E_{\mathrm{\omega}}&=&\ -{{1\over2}\int d^3 r\,
              [g_\omega}j^0_\mathrm{B}({\bf r})\omega_0({\bf r})-g_\omega\omega_\varphi({\bf r}) j^\varphi_\mathrm{B}({\bf r})
               \nonumber \\
              & &\ -\frac{1}{2}c_3(\omega_\mu\omega^\mu)^2({\bf r})]
\end{eqnarray}
and energy contributions from other parts, not shown here, are the
same as given in Ref.~\cite{Yao2006}.


To describe the nuclear magnetic moment, one needs the effective
electromagnetic current which is defined
as~\cite{Furnstahl1989,Yao2006}
\begin{equation}
   \hat{J}^\mu(x) =
                   \bar{\psi}(x)\gamma^\mu\frac{1-\tau_3}{2}\psi(x)+\frac{\kappa}{2M}\partial_\nu
                   [\bar{\psi}(x)\sigma^{\mu\nu}\psi(x)],
\end{equation}
where $\sigma^{\mu\nu}=\frac{i}{2} [\gamma^\mu,\gamma^\nu]$, and
$\kappa$ is the free anomalous gyromagnetic ratio of the nucleon:
$\kappa^p=1.793$ and $\kappa^n=-1.913$. The spatial component of the
effective electromagnetic current operator is given by
\begin{equation}\label{current-operator}
   \mbox{\boldmath{$j$}}(\br)=
               \psi^+(\br)\mbox{\boldmath$\alpha$}\frac{1-\tau_3}{2}\psi(\br)+\frac{\kappa}{2M}\nabla\times
               [\psi^+(\br)\beta\mathbf{\Sigma}\psi(\br)],
\end{equation}
where the first term is the Dirac current, and the second term is
the so-called anomalous current.

From the effective electromagnetic current, the magnetic moment can
be obtained by
\begin{eqnarray}\label{magnetic-moment}
  \mbox{\boldmath{$\mu$}}
            &=& \frac{1}{2}\int d^3r\, \br\times\langle \mathrm{g.s.}|\mbox{\boldmath{$j(r)$}}|\mathrm{g.s.}\rangle\,.  
\end{eqnarray}


\section{Numerical details}

The Dirac equation for nucleons and the Klein¨CGordon equations for
mesons are solved by expansion in terms of the isotropic harmonic
oscillator basis functions in cylindrical coordinates with 14
oscillator shells for both the fermion and the boson
fields~\cite{Gambhir1990,Ring1997}. The oscillator frequency is
given by $\hbar\omega_0=41A^{-1/3}$ MeV. In odd A nuclei, as the
time reversal invariance is broken by the odd nucleon, the Dirac
equation for nucleons should be solved separately in two subspaces
$\{\psi_j\}$ and $\{\psi_{\bar{j}}\}$, where $\psi_j$ and
$\psi_{\bar{j}}$ are time reversal conjugate states~\cite{Yao2006}.

The energy surface is obtained through the deformation constrained
calculation. The quadratic constraint is adopted as in
Ref.~\cite{Ring1980} by constraining the mass quadrupole moment
$\langle\hat{Q}_2\rangle$ to a given value $q$, i.e.,
\begin{equation}
   \langle H'\rangle
   =\langle H\rangle+\frac{1}{2}C(\langle\hat{Q}_2\rangle-q)^2,
\end{equation}
where $\langle H\rangle$ is the total energy, and $C$ is the
stiffness constant. The deformation parameter $\beta_2$ is obtained
from the calculated $\langle\hat{Q}_2\rangle$ for the protons and
neutrons through
\begin{equation}
   \langle\hat{Q}_2\rangle = \langle\hat{Q}_{2p}\rangle  + \langle\hat{Q}_{2n}\rangle
   =\frac{3}{\sqrt{5\pi}}AR_0^2\beta_2,
\end{equation}
with $R_0=1.2A^{1/3}$. In the following, both the adiabatic and the
configuration-fixed deformation constrained
calculation~\cite{Guo2004,Meng2006a,Lu2007} will be performed.

The effective interaction parameter set PK1~\cite{Long2004} is used
throughout the calculation and the center-of-mass (c.m.) correction
is taken into account microscopically by
\begin{equation}
  \label{c.m.}
    E_{\mathrm{c.m.}}^{\mathrm{mic.}}=-\frac{1}{2MA}\langle
    \hat{\mathbf{P}}^2_{\mathrm{c.m.}}\rangle,
\end{equation}
where $\hat{\mathbf{P}}_{\mathrm{c.m.}}$ is the total momentum operator of a nucleus with mass
number $A$.


\section{Results and Discussion}
\begin{figure}
\centering
\includegraphics[width=9cm]{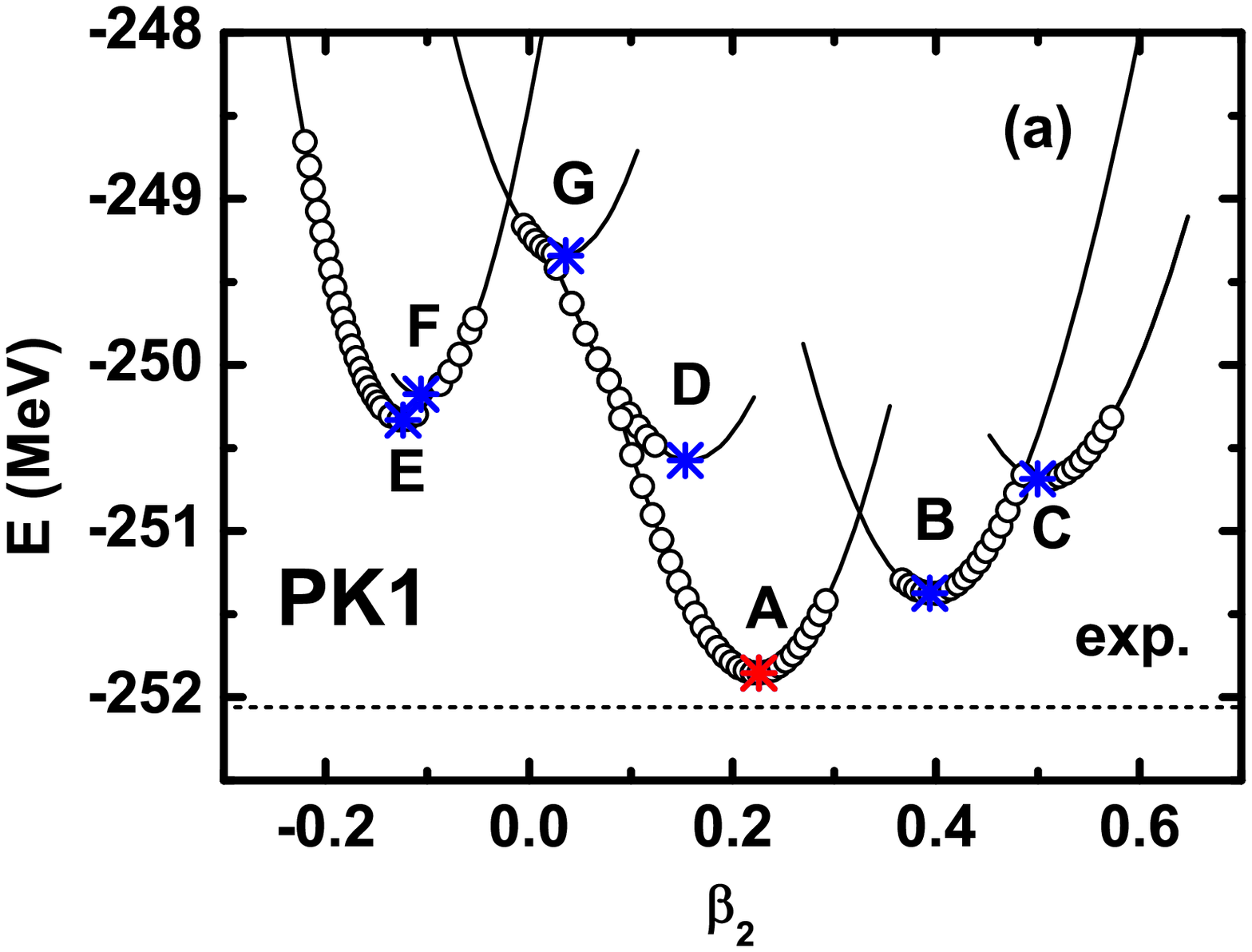}\\
\includegraphics[width=9cm]{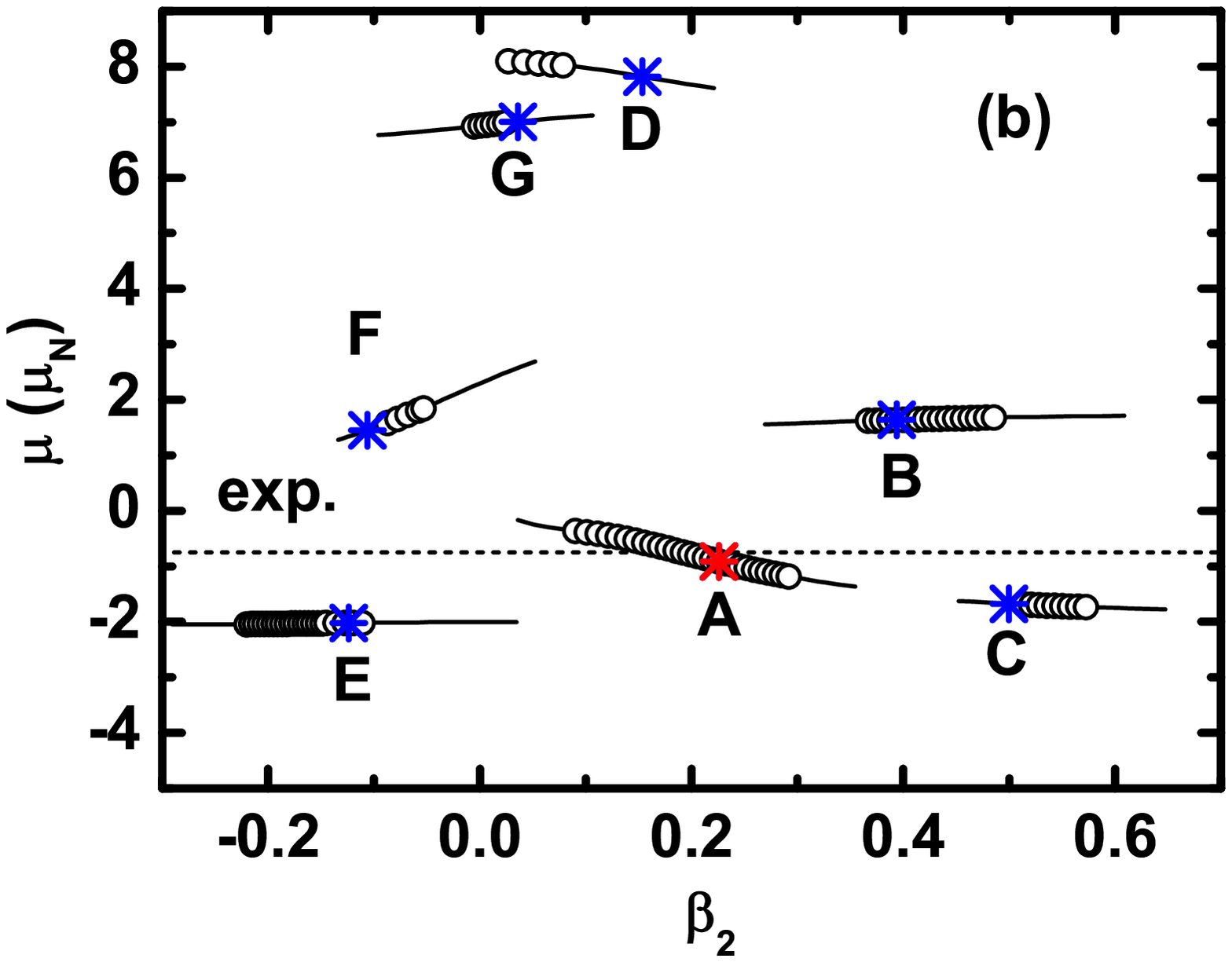}
\caption{\label{fig1} (Color online) (a) The energy surfaces for
$^{33}$Mg as a function of $\beta_2$ by adiabatic (open circles) and
configuration-fixed (solid lines) deformation constrained RMF
approach with time-odd component using PK1 parameter set. The minima
in the energy surfaces for fixed configurations are represented as
stars and respectively, labeled as A, B, C, D, E, F, and G of which
A is the ground state with total energy $E = -251.85$ MeV and
$\beta_2=$ 0.23, in comparison with the data $-252.06$
MeV~\cite{Audi2003} (dotted line). (b) Magnetic moments for the
corresponding configurations in panel (a) as a function of
$\beta_2$. The magnetic moment for the ground state is $
-\,0.913\,\mu_N$ in comparison with the experimental value $ \mu =
-\,0.7456(5)\,\mu_N$~\cite{Yordanov2007} (dotted line). }
\end{figure}
The energy surfaces for $^{33}$Mg as a function of the quadrupole
deformation parameter $\beta_2$ calculated by adiabatic, shown as
open circles, and configuration-fixed, shown as solid lines,
deformation constrained RMF approach with time-odd component using
PK1 parameter set are presented in Fig.~\ref{fig1}(a). As discussed
in the Ref.~\cite{Lu2007}, the configuration-fixed deformation
constrained calculation gives a continuous and smooth curve for the
energy surfaces as a function of $\beta_2$. The local minima in the
energy surfaces for each configuration are represented by stars and
labeled as A¨CG in ascending order of energy. The ground state A is
found to be prolate deformed, $\beta_2=0.23$, with the total energy
$ -251.85$ MeV, which is close to the data $-252.06$
MeV~\cite{Audi2003}. Using Eq.~(\ref{magnetic-moment}), the
effective electromagnetic current gives the nuclear magnetic moment
for given configurations in Fig.~\ref{fig1}(b). It is found that the
magnetic moment is sensitive to the configuration, but not so much
to $\beta_2$. The magnetic moment for the ground state is $
-\,0.913\,\mu_N$ which is in good agreement with the data $ \mu =
-\,0.7456(5)\,\mu_N$~\cite{Yordanov2007}, compared with the
shell-model results $-\,0.675\,\mu_N$ and
$-\,0.705\,\mu_N$~\cite{Yordanov2007} restricted to $2p-2h$
configuration using two different interactions designed specifically
for the island of inversion.

\begin{figure}
\centering
\includegraphics[width=9cm]{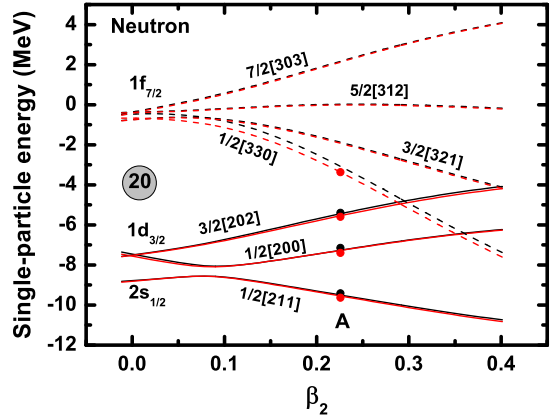}
\caption{\label{fig2} (Color online) Neutron single-particle
energies for $^{33}$Mg as a function of $\beta_2$ obtained by
configuration-fixed deformation constrained calculation for the
configuration of ground state A. Positive (negative) parity states
are marked by solid (dashed) lines. Each pair of time reversal
conjugate states splits up into two levels with the third component
of total angular momentum $\Omega>0$ and $\bar{\Omega}<0$ denoted by
red and black lines respectively. The solid circle denotes that the
corresponding orbitals are occupied in the ground state. }
\end{figure}

In order to examine the evolution of the single particle level and
compare with the results in Ref.~\cite{Yordanov2007}, neutron
single-particle energies for $^{33}$Mg as a function of $\beta_2$
obtained by configuration-fixed deformation constrained calculation
for the configuration of ground state A are presented in
Fig.~\ref{fig2}. The positive (negative) parity states are marked by
solid (dashed) lines, and the occupied orbitals are represented by
filled circles. The self-consistent calculation here gives the odd
neutron in $1/2[330]$ orbital with $\beta_2=0.23$ for the ground
state, while in Ref.~\cite{Yordanov2007}, in a Nilsson-model
picture, the odd neutron in $3/2[321]$ orbital with prolate
deformation $0.3 < \beta_2 < 0.5$ is proposed to reproduce the spin
and parity $I^\pi=\frac{3}{2}^-$.

As the time reversal invariance is broken by the nuclear magnetic
potential, each pair of time reversal conjugate states splits up
into two levels with the third component of total angular momentum
$\Omega>0$ and $\bar{\Omega}<0$. The energy splittings of time
reversal conjugate states range from 0.01 to 0.2 MeV, and the
largest splitting occurs at the orbital occupied by the odd neutron.
At $\beta_2\approx0.3$, the level crossing happens between
$3/2[202]$ and $1/2[330]$ orbitals, and leads to the discontinuity
of adiabatic energy surface at $\beta_2\approx0.3$ in
Fig.~\ref{fig1}(a) as explained in the Ref.~\cite{Lu2007}.
\begin{figure*}
 \centering
 \includegraphics[width=14cm]{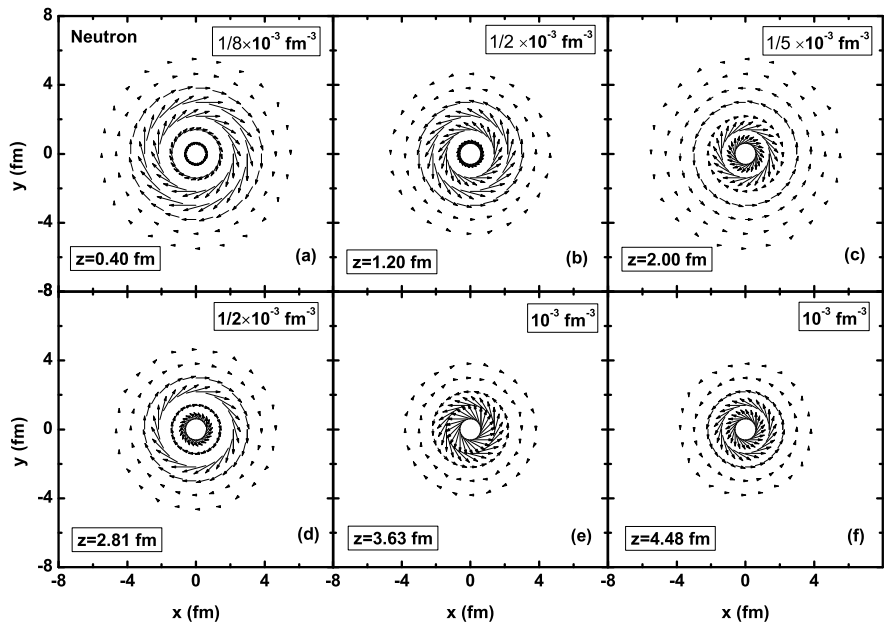}\\
 \vspace*{-1cm}
 \caption{\label{fig3}
 Azimuthal components of neutron current in the $xy$ plane at $z = 0.40$ fm (a),
 $z = 1.20$ fm (b), $z = 2.0$ fm (c), $z = 2.81$ fm (d), $z = 3.63$ fm (e), and $z = 4.48$ fm (f)
 respectively for the ground state of $^{33}$Mg. The direction and length of the arrows
 respectively represent the orientation and magnitude of current which is in different
 scale for different panels.}
\end{figure*}

In Fig.~\ref{fig3}, as an illustration, the neutron current in $xy$
planes is plotted at $z = 0.40 \sim 4.48$ fm for the ground state of
$^{33}$Mg. The direction and length of the arrows respectively
represent the orientation and magnitude of current, which is in
different scale for different panels. Take Fig.~\ref{fig3}(a) as an
example, since only azimuthal components of neutron current exist
under the axial symmetry, the neutron current is on circular lines
around the symmetry axis and peaks at $r_\perp=3\sim4\,\mathrm{fm}$.
From panel (a) to (f), one can see that, as $z$ increases, the
neutron current gradually grows larger and there is a tendency of
closing up to the center in the $xy$ plane. But in general, it peaks
at $r\approx R_n=3.4$ fm (the neutron rms radius), indicating that
the current is mainly contributed from the old nucleon and flows
around the nuclear surface~\cite{Yao2006}. It is interesting to note
that for fixed $z$, the neutron current can be clock-wise,
anti-clock-wise or even
a mixture. 

\begin{figure}
 \centering
 \includegraphics[width=9cm]{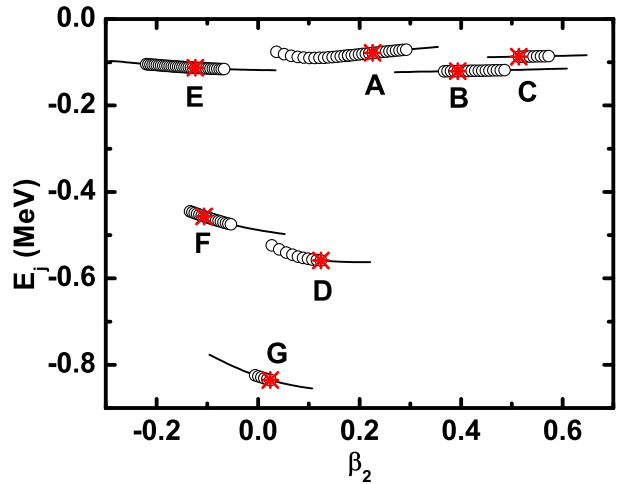}
 \caption{\label{fig4} (Color online)
 The energy contributions due to the nuclear magnetic potential, $E_\textrm{j}\,=-1/2\int
 d^3r\,g_\omega\omega_\varphi({\bf r}) j^\varphi_\mathrm{B}({\bf r})$,
 as a function of $\beta_2$
 for different configurations.}
\end{figure}

In order to investigate the effect of the nuclear magnetic potential
on the binding energy, in Fig.~\ref{fig4} the energy contributions
for different configurations as a function of $\beta_2$ are given.
In general, the nuclear magnetic potential makes the nucleus more
tightly bound, and numerically, its contribution to energy is
several hundred keV. Specially, the contributions for A, B, C, E are
around $-0.1$ MeV, while for D, F, G $-0.5\sim-0.8$ MeV, which is
due to the breaking of proton pair shown in the following.
\begin{table*}[]
\tabcolsep=3pt \caption{The energy ($E_{\textrm{cal}}$), the
quadrupole deformation parameters ($\beta_{\textrm{cal}}$), the
valence nucleon configuration, the magnetic moments of the nucleus
($\mu_{\textrm{total}}$) and the core $(\mu_{\textrm{core}})$, in
comparison with experimental value of the ground-state energy
($E_{\textrm{exp}}$)~\cite{Audi2003} and magnetic moment
($\mu_{\textrm{exp}}$)~\cite{Yordanov2007}. The energy is in MeV and
the magnetic moment in $\mu_{\mathrm{N}}$. }
\begin{tabular}{ccccccccc}
  \hline\hline
  ~State~ & $~~E_{\textrm{cal}}(E_{\textrm{exp}})$~~ &~~~$\beta_{\textrm{cal}}$~~~& ~~Valence nucleon configuration~~~
  &~$\mu_{\textrm{total}}(\mu_{\textrm{exp}})$~ &~~$\mu_{\textrm{core}}$~~ \\ \hline
     A    & -251.85(-252.06)   &  0.23   & $\nu\frac{1}{2}[330-]$                                         &-0.9134(-0.7456)&-0.042   \\
     B    & -251.37   & ~0.40   & $\nu\frac{3}{2}[202+]$                                                  &~1.6380         &-0.127   \\
     C    & -250.69   & ~0.50   & $\nu\frac{3}{2}[321-]$                                                  &-1.6750         &-0.071   \\
     D    & -250.58   & ~0.15   & $\nu\frac{1}{2}[330-]\otimes\pi\{\frac{3}{2}[211+]\frac{5}{2}[202+]\}$  &~7.8179         &-0.216   \\
     E    & -250.33   & -0.12   & $\nu\frac{7}{2}[303-]$                                                  &-2.0198         &-0.114   \\
     F    & -250.18   & -0.11   & $\nu\frac{7}{2}[303-]\otimes\pi\{\frac{1}{2}[211+]\frac{3}{2}[211+]\}$  &~1.4490         &-0.239    \\
     G    & -249.34   & ~0.04   & $\nu\frac{5}{2}[312-]\otimes\pi\{\frac{3}{2}[211+]\frac{5}{2}[202+]\}$  &~7.0082         &-0.288   \\ \hline\hline
\end{tabular}\label{table1}
\end{table*}

The total energy ($E_{\textrm{cal}}$), quadrupole deformation
parameter ($\beta_{\textrm{cal}}$), valence nucleon configuration,
the magnetic moments of the nucleus ($\mu_{\textrm{total}}$) and
core $(\mu_{\textrm{core}})$ for different configurations are listed
in Table.~\ref{table1}. It is obvious that the magnetic moments are
sensitive to the valence nucleon configurations. For states A, C and
E, the magnetic moments are negative, because the main component of
corresponding valence nucleon wave function belongs to
$\nu\,1f_{7/2}$ which has a negative Schmidt value. For state B, the
main component of valence nucleon wave function belongs to
$\nu\,1d_{3/2}$, and thus gives a positive magnetic moment. For
states D and G, the total magnetic moments are remarkably large due
to the breaking of proton pair and the enhanced Dirac current. For
F, although there are three valence nucleons, the total magnetic
moment is smaller than states D and G due to the strong cancelation
between the valence protons and neutron.

The core polarization effect can be investigated by examining the
magnetic moments of the core $(\mu_{\textrm{core}})$. The baryon
current of the valence nucleons is responsible for the polarization
which results in the nonvanishing $\mu_{\textrm{core}}$. As it is
seen from the Table.~\ref{table1}, the $\mu_{\textrm{core}}$, which
is caused by the polarization currents, is negative for all states.
It is obvious that $\mu_{\textrm{core}}$ for states D, F and G are
much larger than states A£¬B£¬C and E, due to the polarization
effect of more valence nucleons.

\section{Summary}
In summary, the configuration-fixed deformation constrained RMF
approach with time-odd component has been applied to investigate the
ground-state properties of $^{33}$Mg with effective interaction PK1.
Using the configuration-fixed deformation constrained calculation,
the ground state of $^{33}$Mg has been found to be prolate deformed,
$\beta_2=0.23$, with the odd neutron in $1/2[330]$ orbital. The
ground-state energy $ -251.85$ MeV is close to the experimental
value $-252.06$ MeV~\cite{Audi2003}. Using
Eq.~(\ref{magnetic-moment}), the magnetic moment
$-\,0.913\,\mu_\mathrm{N}$ is obtained with the effective
electromagnetic current which well reproduces the data $
-\,0.7456\,\mu_\mathrm{N}$~\cite{Yordanov2007} self-consistently
without introducing any parameter, in contrast with the shell-model
results $-\,0.675\,\mu_\mathrm{N}$ and
$-\,0.705\,\mu_\mathrm{N}$~\cite{Yordanov2007} restricted to $2p-2h$
configuration using two different interactions designed specifically
for the island of inversion. The energy splittings of time reversal
conjugate states, the nucleon current, the energy contribution, and
the effect of core polarization due to the nuclear magnetic
potential are discussed in detail.

Apart from the core polarization, the meson exchange current
correction is also very important for the descriptions of nuclear
magnetic moment, especially the isovector magnetic moment.
Investigations along this line are in progress.


\begin{acknowledgments}
This work is partly supported by Major State Basic Research
Developing Program 2007CB815000 as well as the National Natural
Science Foundation of China under Grant Nos. 10775004, 10221003,
10720003 and 10705004.
\end{acknowledgments}


\begin{thebibliography}{90}
 \bibitem{Blin-Stoyle1956}
 Blin-Stoyle R J. Theories of Nuclear Moments. Rev Mod Phys, 1956, 28: 75-101

 \bibitem{Arima1984}
 Arima A. Nuclear magnetic properties and gamow-teller transitions. Prog Part Nucl Phys, 1984, 11:
 53-89

 \bibitem{Castel1990}
 Castel B, Towner I S. Modern Theories of Nuclear Moments. Oxford: Clarendon Press,
 1990.

 \bibitem{Stone2005}
 Stone N. Table of nuclear magnetic dipole and electric quadrupole moments. At Data Nucl Data Tables, 2005, 90:
 75-176

 \bibitem{Neyens2003}
 Neyens G. Nuclear magnetic and quadrupole moments for nuclear structure research on exotic nuclei. Rep Prog Phys, 2003, 66:
 633-689

 \bibitem{Warburton1990}
 Warburton E K, Becker J A, Brown B A. Mass systematics for A=29-44 nuclei: The deformed A~32 region. Phys Rev C, 1990, 41:
 1147-1166

 \bibitem{Nummela2001}
 Nummela S, Nowacki F, Baumann P, et al. Intruder features in the island of inversion: The case of $^{33}$Mg. Phys Rev C, 2001, 64: 054313

 \bibitem{Pritychenko2002}
 Pritychenko B V, Glasmacher T, Cottle P D, et al. Structure of the "island of inversion" nucleus $^{33}$Mg. Phys Rev C, 2002, 65: 061304

 \bibitem{Elekes2006}
 Elekes Z, Dombradi Zs, Saito A, et al. Proton inelastic scattering studies at the borders of the "island of inversion": The $^{30,31}$Na
 and $^{33,34}$Mg case. Phys Rev C, 2006, 73: 044314

 \bibitem{Tripathi2008}
 Tripathi V, Tabor S L, Mantica P F, et al. Intruder Configurations in the A = 33 Isobars: $^{33}$Mg and $^{33}$Al. Phys Rev Lett, 2008, 101: 142504

 \bibitem{Yordanov2007}
 Yordanov D T, Kowalska M, Blaum K, et al. Spin and Magnetic Moment of $^{33}$Mg: Evidence for a Negative-Parity Intruder Ground State.
 Phys Rev Lett, 2007, 99: 212501.

 \bibitem{Lawson1980}
 R. D. Lawson. Theory of Nuclear Shell Structure. Oxford: Clarendon Press,
 1980. 300

 \bibitem{SEROT1986}
 Serot B D, Walecka J D. The Relativistic Nuclear Many-body Problem. Adv Nucl Phys, 1986, 16:
 1-327

 \bibitem{Ring1996}
 Ring P. Relativistic mean field theory in finite nuclei. Prog Part Nucl Phys, 1996, 37:
 193-263

 \bibitem{Vretenar2005}
 Vretenar D, Afanasjev A, Lalazissis G, et al. Relativistic Hartree-Bogoliubov theory: static and dynamic aspects of exotic nuclear structure.
 Phys Rep, 2005, 409: 101-259

 \bibitem{Meng2006}
 Meng J, Toki H, Zhou S, et al. Relativistic continuum Hartree Bogoliubov theory for ground-state properties of exotic nuclei.
 Prog Part Nucl Phys, 2006, 57: 470-563

 \bibitem{Hofmann1988}
 Hofmann U, Ring P. A new method to calculate magnetic moments in relativistic mean field theories.
 Phys Lett B, 1988, 214: 307-311

 \bibitem{Furnstahl1989}
 Furnstahl R J, Price C E. Relativistic Hartree calculations of odd-A nuclei. Phys Rev C, 1989, 40:
 1398-1413

 \bibitem{Yao2006}
 Yao J M, Chen H, Meng J. Time-odd triaxial relativistic mean field approach for nuclear magnetic moments.
 Phys Rev C, 2006, 74: 024307

 \bibitem{Konig1993}
 K$\ddot{o}$nig J, Ring P. Phys Rev Lett, 1993, 71: 3079-3082

 \bibitem{Madokoro2000}
 H. Madokoro, J. Meng, M. Matsuzaki, et al. Relativistic mean field description for the shears band mechanism in $^{84}$Rb.
 Phys Rev C, 2000, 62: 061301

 \bibitem{Peng2008a}
 Peng J, Meng J, Ring P, et al. Covariant density functional theory for magnetic rotation.
 Phys Rev C, 2008, 78: 024313

 \bibitem{Li2009}
 Li Jian, Yao J M, Meng J. Deformation constrained relativistic
 mean-field approach with fixed configuration and time-odd component.
 Chin Phys C, 2009, 33(S1): 98

 \bibitem{Meng2006a}
 Meng J, Peng J, Zhang S Q, et al. Possible existence of multiple chiral doublets in $^{106}$Rh. Phys Rev C, 2006, 73: 037303

 \bibitem{Reinhard1989}
 Reinhard P G. The relativistic mean-field description of nuclei and nuclear dynamics. Rep Prog Phys, 1989, 52: 439-514

 \bibitem{Gambhir1990}
 Gambhir Y K, Ring P, Thimet A. Relativistic mean field theory for finite nuclei. Ann Phys, 1990, 198: 132-179

 \bibitem{Ring1997}
 Ring P, Gambhir Y K, Lalazissis G A. Computer program for the relativistic mean field description of the ground state properties of even-even axially deformed nuclei.
 Comput Phys Commun, 1997, 105: 77-97

 \bibitem{Ring1980}
 Ring P, Shuck P. Nuclear Many Body Problem. New York: Springer Press,
 1980. 266-271

 \bibitem{Guo2004}
 Guo L, Sakata F, Zhao E.-G. Characteristic feature of self-consistent mean-field in level crossing region.
 Nucl Phys A, 2004, 740: 59-76

 \bibitem{Lu2007}
 L$\ddot{u}$ H, Geng L S, Meng J. Constrained relativistic mean-field approach with fixed configurations.
 Eur Phys J A, 2007, 31: 27

 \bibitem{Long2004}
 Long W, Meng J, Giai N V, et al. New effective interactions in relativistic mean field theory with nonlinear terms and density-dependent meson-nucleon coupling.
 Phys Rev C, 2004, 69: 034319

 \bibitem{Audi2003}
 Audi G, Wapstra A H, Thibault C. The 2003 atomic mass evaluation: (II). Tables, graphs and references. Nucl Phys A, 2003, 729: 337-676
\end{thebibliography}
 
\end{document}